# Low-dose ionizing radiation exposure during pregnancy induces behavioral impairment and lower weight gain in adult rats

GIAROLA, R.S.[1*]; DE ALMEIDA, G.H.O.[1]; HUNGARO, T.H.[1]; DE PAULA, I.R.[1]; GODINHO, A.F.[2]; ACENCIO, M.L.[3]; MESA, J.[1]; LEMKE, N.[1]; VIEIRA L.D.[1]; DELICIO, H.C.[4]

[1]Department of Physics and Biophysics, Bioscience Institute, São Paulo State University (UNESP), Botucatu, Brazil.
[2]Center for Toxicological Assessment (CEATOX), Bioscience Institute, São Paulo State University (UNESP), Botucatu, Brazil.
[3]Department of Clinical and Molecular Medicine, Norwegian University of Science and Technology. Trondheim, Norway.
[4]Department of Physiology, Bioscience Institute, São Paulo State University (UNESP), Botucatu, Brazil.
*Correspondence to chancez@hotmail.com

## Abstract

Low-dose ionizing radiation may induce far-reaching consequences in human, especially regarding intrauterine development. Many studies have documented that the risks of in utero irradiation remain controversial and no effect is reported at doses below 50 mGy. Animal models are often used to clarify the non-fully understood impact of intrauterine irradiation and allow the manipulation of several experimental setups, making possible the analysis of a wide range of end points. We investigated the impact of in utero low-dose X-ray irradiation on postnatal development in rat offspring through a set of well-established behavioral parameters and weight gain. To investigate the hypothesis of postnatal behavioral and physiological alterations due to prenatal low-dose ionizing radiation we exposed pregnant Wistar to 15 mGy of X-rays on gestational days 8 and 15 and control mothers. This low-dose value into diagnostic range can be achieved in a single radiological exam. Four male animals were select from each litter. At infant age, eye-opening test and negative geotaxis tests were performed. Animals were tested at postnatal ages 30 and 70 days in open field, elevated plus-maze, and hole board tests. We evaluated the weight gain of all animals throughout the experiment. The results presented differences between irradiated and non-irradiated animals. Exposed animals presented lower weight gain in adult life, impairment in central nervous system since infant phase, behavioral alterations persisting into later life, and motor coordination impairment. Effects at doses under 100 mGy have not been reported, however, the present study demonstrate that 15 mGy intrauterine exposure was able to generate deleterious effects.

## Introduction

It is widely accepted that ionizing forms of diagnostic imaging should not be used in pregnant women due to the potential harmful biological effects of ionizing radiation (IR) on the embryo and fetus [1]. The IR can produce direct ionization in the DNA affecting its functionality or even destroying the DNA, IR can also produce indirect effects by the energy deposited in a biological environment and then trigger the production of reactive free radicals which, in turn, increase cellular oxidative stress [1]. Oxidative stress may cause cellular phenotypic changes that can predispose the fetus to certain post-natal phenotypes [2], negative effects are usually expected due to IR-induced oxidative stress [3]. Therefore, to avoid these possible negative effects, it is preferable not to expose embryo and fetus tissues to IR. However, ionizing radiation induce negative effects at all exposure levels in humans?

While the biological effects associated with prenatal exposure to high-dose (> 100mGy) ionizing radiation in humans have been widely investigated [4], studies reporting the biological effects of low-dose (≤ 100 mGy) ionizing radiation (LDIR) are scarce and the interpretation of its results is often controversial [1,5-10]. A pioneering study of childhood cancers following irradiation during fetal life was conducted by Dr. Alice Stewart in 1956 [7,10]. This first report inspired further research to conduct studies aiming to establish a strong correlation between LDIR from diagnostic imaging procedures during pregnancy and adverse health effects on growth and development. For example, D'ippolito and Bitelli [1] reported that exposure to radiation doses below 50 mGy is not associated with increased risk of miscarriage, congenital anomalies, mental retardation or neonatal mortality. In addition, Cohen et al. [11] studying pregnant women exposed to radiographic diagnostic procedures during pregnancy, showed that the rate of live births and miscarriages was, respectively, higher and lower in the low-dose group (< 1.0 mGy during pregnancy). In addition, they argued that ionizing radiation has its effect directly on the fetus, differing from drugs, so there are no concerns about differences in metabolism, absorption, or placental transfer between the animal and the human model. However, the Cohen et al. report concludes that the threshold for effects such as growth retardation and teratogenesis is 200 mGy [11], while National Council on Radiation Report (NCRP) No. 174 established the threshold level of 50 mGy [12]. No solid correlation could be established from the available literature that LDIR results in an increased risk of stochastic and deterministic effects to the offspring [5,8]. Therefore, there is no currently solid evidence based on well-controlled studies that LDIR induces deleterious effects in humans.

As it is not possible to perform controlled experiments to investigate the effects of LDIR in humans for ethical and moral reasons, model organisms, such as rats, should be used instead. Rodent models allow precise manipulation of all influencing factors [5] and provide fundamental biological and behavioral data. Furthermore, due to it the high reproductive rate, rapid body development, easy manipulation, good learning ability, and a shorter life span compared to humans, it is possible to investigate long-term effects [5]. The correspondence between the embryonic stages of rat and human is described by Hill [13], making it possible to carry the observations made in rats to the human in relation to the irradiation period.

Exposure to LDIR during pregnancy can lead to behavioral alterations, and animals exposed in utero may exhibit behavioral changes during maturation [14]. Some studies recognize brain damage in the post-natal period [15], the available literature describes harmful effects on the developing cerebral cortex and altered post-natal behavior [6,16]. Several studies have been conducted on the irradiation of pregnant rats, but generally, the radiodiagnostic range (<50 mGy) is not related to the increased risk of congenital anomalies [17,18]. De Santis et al. [17] discuss that doses used in animal studies are generally higher than doses used in diagnostic procedures in humans, therefore, the present study aims to search for effects at low-doses. Behavioral tests and neurodevelopmental reflexes are tools for determining changes that could have been caused by brain injury, and abnormalities in the developmental of the offspring [19]. Otake et al. reported that children exposed in utero to the Hiroshima atomic bomb were clinically identified to be mentally retarded with reduced intelligence quotient (IQ) scores [8,20].

The experimental models of behavioral assessment selected in the present study are widely used in risk assessment and evidence damage to the central nervous system. Rodents, such as the Wistar rat, exhibit a natural aversion to lighted areas and a characteristic behavior that leads them to remain close to vertical surfaces (thigmotaxis), these animals also tend to exploit the threatening stimuli (neophilia). The result of these conflicting impulses is anxiety. The decrease of anxiety leads to an increase in exploratory behavior and, in an antagonistic way, the increase of anxiety leads to lesser locomotion and the tendency to remain at the edges of the field [21]. Thus, the tests allow evaluating the behavior of rats in relation to anxiety in terms of their abilities to acquire and use spatial information as well as habituation to exploratory activity over time.

We examined the effects of in utero ionizing radiation exposure on postnatal physiological, behavioral and locomotor outcomes in infant, young (30 days old) and adult (70 days old) Wistar rats. Pregnant Wistar were exposed to 15 mGy in two distinct stages of development: organogenesis (gestational day 8 [GD8], considered the onset of the organogenesis phase) and fetal development (gestational day 15 [GD15], considered the end of organogenesis phase) [15]. The teratogenic effects of LDIR during organogenesis can lead to intrauterine growth restriction (IURG) [17], such as lethal effects, malformations, structural alterations, mental retardation and cancer induction [12,22]. As pregnancy has its inherently associated risks, it is difficult to measure the effects of LDIR during intrauterine exposure. Therefore, the present study was designed to avoid any external interference that could negatively influence pregnancy in rats.

## Materials and Methods

### Animals and breeding

Adult female (F0, 85 days of age, 150g, *n* = 36) and male (85 days of age, 230g, *n* = 15) virgin albino Wistar rats were obtained from breeding colony housed at the Sao Paulo State University (UNESP). Twenty days before the beginning of the experiment the rats were housed in groups of five in polypropylene cages (40 x 50 x 20 cm) at controlled room temperature (23 ± 2°C), on a 12 h light/dark cycle starting at 6 a.m., standard rat pellet chow (Presence@, Brazil) and tap water were available *ad libitum* during the whole experiment.

Matings were conducted during overnight, by placing a male breeder in female housing, on the next morning females were checked for evidence. Mating was confirmed by the presence of spermatozoa in the contents of the vaginal smear. After pregnancy confirmation, females were housed singly, and this moment was designated as gestational day 1 (GD1). All animals were weighed every three days.

### Statement

All experiments and methods were performed in accordance with relevant guidelines and regulations. This research agrees with the Ethical Principles in Animal Research adopted by Brazil's National Council for the Control of Animal Experimentation. All procedures were approved by the Committee of Ethics on the Use of Animals (887/2016-CEUA), Institutional Ethical Committee of UNESP at Botucatu, Brazil.

### Irradiation and postnatal follow-up

The pregnant rats were randomly divided into three experimental groups (*n* = 36): sham-irradiated control (CTL, *n* = 12), irradiated on the eighth day of pregnancy (IR8d, *n* = 12) and irradiated on the fifteenth day of pregnancy (IR15d, *n* = 12). The pregnant Wistar were whole-body irradiated in a circular lucite chamber with 15 mGy of X-rays, with a dose rate of 18 mGy per second, this value was defined because is it a low-dose value into diagnostic range that can be achieved in a single radiological exam and is within the reference dose levels adult abdomen exam [23].

The irradiation was performed by a Shimadzu EZy-Rad Radiodiagnostic machine operating at 81 kVp and 110 mAs, with single 1.2 mm focus, 1.5 mm Al of permanent filtration plus 1.0 mm Al collimator filtration (focus tissue distance = 83 cm). The sham-irradiated control group underwent a similar experimental procedure to those irradiated, except that the machine was turned off. A calibrated Radcal Corporation ionization chamber (model 90x5-6) connected to a Radcal Corporation electrometer (model 9010) was used to determine the dose delivery parameters (15 mGy), the equipment was placed inside the circular lucite chamber at the central position where the rats were irradiated. After a whole-body single-dose exposure to X-rays, the F0 body weights were measured every day until the birth of the offspring.

Dams were allowed to give birth and nurture their offspring (F1) normally, and on the day after delivery all the litters were examined externally and sexed. Litters were culled to four males in each litter on post-natal day 3 [24]. Weaning was performed when animals were 21 days old. All animals were marked twice a week with colored pens to track all the physical and behavioral developmental evaluations. F1 animals were weighed individually every two days from postnatal day 1 to day 91.

**Assessment of reflex development and physiological and behavioral responses**

The reflex development was evaluated by eye-opening and negative geotaxis tests in infant rats. The physiological response evaluated in the present study was weight gain. The anxiety-related behavior responses were evaluated by open field, elevated plus maze and hole board tests. Rat behavioral evaluations were performed, firstly at a young age, with 30 days of life, and secondly at an adult age, with 70 days of life, using open field, elevated plus maze and hole board tests. Animals were tested without prior habituation in order to avoid adaptation since these tests should capture the natural conflict between the exploration and the tendency to avoid new environments and behavioral responses can only be observed before habituation occurs [25,26]. The experiments were conducted between May 2016 and January 2017.

*Eye-opening test*

Eye-opening (EO) test indicates abnormalities in neurodevelopmental reflex in neonatal rodents. All pups were checked daily since birth, and the number of days that the four pups of the same litter opened their eyes was considered [19].

*Negative geotaxis test*

Negative geotaxis (NG) test is used for sensorimotor development during the early days of post-natal development. Beginning on day 3 after the birth, the pups were individually positioned face down on a 45° inclined surface ramp and the time required for the pup to rotate at least 90°, not exceeding 60 seconds per trial, was recorded. Animals that were unable to perform the task were counted as unrealized until they were able. The first climb was considered the number of days in which all the pups of the same litter in each group turned around at least 90° [26].

**Behavioral tests**

*General considerations and experimental conditions*

The room for behavioral assessment was sound-proof, temperature-controlled and illuminated by dim red lights. The period of behavioral observations was defined between 9 a.m. and 11 a.m. The animals were submitted to three behavioral tests to indicate anxiety, in the following order and sequentially: the open field (OF), elevated plus-maze (EPM) and hole board (HB). All these tests indirectly measure anxiety by measuring exploratory and locomotor activities. We opted to perform the three tests to mitigate misinterpretation of the results obtained by an individual test [28,29]. Usually, a combination of tests is used to improve the analysis of the animal's anxiety, as it is a transient condition that is only noticed at given moment [26]. All the behavioral tests were performed on the same day, during two phases: when the animals were 30 and 70 days of age, all pups in each litter (F1 animals) underwent the subsequential behavioral tests. Animals that fell on the elevated plus-maze test had their data disregarded for further analysis, but the number of falls was counted. To prevent observational bias in behavioral assessment, the three testers were blind to the treatment group. The apparatus used was thoroughly cleaned with a 5% ethanol solution before each animal was introduced. The maximum care in handling was taken to not influence the responses obtained, the animals were familiarized with manipulation of the handlers. The proper handling may avoid changes of mood, discomfort, and increasing excitation, factors that may influence the outcomes, including rearing behavior [30].

*Open field test*

The open field (OF) test is an experiment used to analyze the levels of general locomotor activity, emotionality and exploratory behavior in rodents used for scientific research [31]. The test device consists of a round wooden arena (100 cm in diameter) surrounded by walls (32 cm high walls) that prevent the animal from escaping. The round area is demarcated by painted black stripes into three concentric circles: central, medium and external (diameters of 18, 57 and 100 cm, respectively), and there are also subdivision quadrants painted in which circle: 6, 6 and 12, respectively, these demarcations serve to account for the movement of the animal. The tested animal is subjected to an unknown new environment from which it is not able to escape.

During open field observations, each rat was individually placed in the center of the arena and for the next three minutes, the rat movements were recorded by a camera mounted above the arena. Through the analysis of the recorded videos, locomotor activity (number of demarcated spaces that the animal entered with the four paws), rearing (the frequency animals rise up on its hind limbs), time spent to cross the center for the first time (the time spent for the first time that the animal crossed the center circle of the arena), and the number of crossings by the arena central circle (number of times that the animal crossed the arena center) were counted. The Video 1 present a part of the recorded test and Figure 1 presents the Open Field test device.

*Elevated plus-maze test*

The elevated plus-maze (EPM) is a test used to investigate rodent anxiety [32]. The apparatus consists of two open arms crossed with two closed arms of equal length and width (50 x 10 cm each) and the center – the place where the arms join (10 x 10 cm). The EPM is 50 cm elevated above the floor and the open arms have a plexiglass edge 1 cm high while the enclosed arms have wooden walls that extend 40 cm high. This test is based on the principle that the rat can freely explore the open arms but may also remain protected (thigmotaxis) in the closed arms, normally rats tend to avoid open areas. This conflict allows to measure anxiety or fear-induced decrease in normal exploratory activity; in addition, the number of entries in both arms is used as a measure of locomotor activity and exploratory behavior [33]. Anxiolytic drugs led the rats to spend more time on open arms, while anxiogenic does the opposite [34].

Each animal was individually placed in the center of the EPM facing a closed arm and allowed to explore freely for five minutes. Total entries with all the paws and the time that each animal spent exploring (seconds) were counted through the analysis of recorded videos using a camera mounted above the apparatus [35]. Sometimes animals can fall off the device's open arms, although it is not common to happen. We recorded these events, but we disregarded them for further analyses because the anxiety levels of animals that experienced falls were likely to be affected, and therefore the results of the following test, the hole board test, for these animals could not be reliable [36]. Figure 2 presents the Elevated Plus Maze test device. The Video 2 present a part of the recorded test and Video 3 present a fall.

*Hole Board*

The Hole Board (HB) test is an experimental method that measures anxiety, exploratory behavior, and locomotor activity. The device is a square wooden arena with 57 cm of edges (floor) and a surrounding wall of 30 cm high, the floor has 81 (9 x 9) round holes (2.5 cm diameter) distributed throughout the floor and separated from each other by 3.5 cm [37]. The rats were individually placed in the center of the apparatus and, for the next 5 minutes, the movement was recorded by a camera placed under the device. During the test the rats could move freely. The number of head and legs dip were counted. The rat's curiosity causes it to poke his nose into a hole, so the head dip was only considered when both eyes were in the low level of the hole. The leg dip was only counted when all the fingers passed through the hole [38]. Figure 3 presents the Hole Board test device and Video 4 present a part of the recorded test.

**Statistical analysis**

To determine the significance of physiological and behavioral differences between the irradiated and control groups, we used One-way Analysis of Variance (ANOVA) in conjunction with Tukey's honestly significant difference (HSD) post-hoc. All statistical analyzes were performed using the STATISTICA software [39]. Results were considered significant at p-values < 0.05.

# Results

**Alterations in gestational features due to irradiation**

In our study, following irradiation with 15 mGy of X-ray on gestational days 8 and 15, irradiated pregnant rats F0 weight gain did not differ from unirradiated control group. Pregnancy duration and litter size were also not significantly different between irradiated and control groups ($P > 0.7$ for pregnancy duration and $P > 0.3$ for litter size). Moreover, mortality was not observed in any group. The body weight, pregnancy duration, and litter size are given in Table 1. The mean body weights of F0 rats four days before and three days after the parturition are illustrated in Figure 4.

**Effects of intrauterine exposure on postnatal body weight**

There was no significant difference between the irradiated and control pups body weights (F1 animals at 96 hours of age) ($P > 0.75$) and between body weights of irradiated and control F1 animals aged 30 days; however, body weights of F1 animals near adult life were significantly different ($P < 0.01$) between control and irradiated animals (irradiated on gestational days 8 [IR8d] and 15 [IR15d]). Irradiated animals had lower weight gain when compared to control, the mean body weights and statistical values are given in Table 2. Figure 5 illustrated F1 rat mean body throughout the whole experiment and Figure 6 illustrated F1 rat mean body weight from day 70 to 91.

**Early sensorimotor development is impaired by in utero LDIR**

To check whether prenatal LDIR was able to permanently alter the responsiveness of the postnatal nervous system, we first assessed the early sensorimotor development. To this end, the results obtained by the eye-opening and negative geotaxis tests were analyzed. Figure 7 presents one F1 newborn animal in the negative geotaxis tests, and a sequence of photos in which the animal performed the turn in the negative geotaxis test.

In the eye-opening test, we evaluated the mean number of days each group opened their eyes after birth. IR8d animals, but not IR15d animals, delayed eye-opening significantly in relation to CTL animals ($P < 0.01$), the values are given in Table 3. This was the first indicative factor that prenatal irradiation, specifically on GD8, could impair sensorimotor development and prenatal nervous system responsiveness.

In the negative geotaxis test, we analyzed the mean number of days each group performed its first climb after birth. We observed that IR8d and IR15d animals made their first climbs significantly later ($P < 0.05$) than CTL animals, the values are given in Table 3. Markedly, IR8d animals presented higher latency for the first climb. However, the mean time to climb did not present statistical difference.

**In utero LDIR at gestation day 8 generates anxiety in rats**

The impairment of sensorimotor development by in utero LDIR, especially LDIR on the GD8, is an evidence that postnatal nervous system responsiveness is altered due to prenatal LDIR. To reinforce this, we sought to investigate whether prenatal exposure to LDIR could also generate postnatal anxiety.

We firstly performed the OF test and four behavioral parameters were analyzed: the number of squares crossed, rearing frequency, time to first cross the center and number of crossings by the center – indicators of locomotor and exploratory activities [40]. The number of squares crossed by young IR8d rats (30 days old rats) was significantly higher than the crossed by young CTL rats ($P < 0.01$) and no significant difference could be detected when comparing young IR15d rats to CTL ones, summarized in Table 4. For adult rats (70 days old), on the other hand, the number of squares crossed by IR8d and IR15d rats were significantly lower ($P < 0.01$) than those crossed by adult CTL rats of the same age. The rearing frequency was also significantly different between the irradiated groups and the control group: IR8d animals presented a higher rearing frequency than the control group ($P < 0.01$) and, contrarily, IR15d animals had a lower rearing behavior than the control group ($P < 0.01$) (Table 4). Regarding the time to first cross the center and the number of crossings by the center, there were no statistically significant differences for both adults and young rats when we compared irradiated with sham-irradiated control animals (Table 4), suggesting that these two parameters did not detect changes in exploratory activities in the present study.

We evaluated next anxiety by the EPM test [32]. In this test, high number of entries in closed arms associated with a long time spent in these arms indicate anxiety [28]. The evaluation of the EPM test is summarized for irradiated and sham-irradiated control young and adult rats in Table 5. Young IR8d animals entered on the open arms significantly more times than CTL mice ($P < 0.01$) while adult IR8d rats entered significantly ($P < 0.01$) less times than CTL animals. Regarding the number of entries into open arms, young IR8d rats entered significantly more times into open arms than CTL rats ($P < 0.01$) and adults IR8d rats entered statistically significant fewer times into open arms than CTL animals (Table 5). When we compare the number of entries into closed arms, we can observe significant differences ($P < 0.01$) only for adult IR15d rats: these animals entered significantly fewer times into closed arms than CTL animals (Table 5). Relevant data were obtained with the number of animals that fell off the device's open arms, none of the young animals fell off the open arms, while for adult control and IR15d one animal fell off for each group, and IR8d eight animals fell off from open arms. Another way to assess the data of EPM test is by calculating the percentage of time on open arms, percentage of entries in open arms and the total number of entries in both arms, this data is shown in Table 6 [34].

Finally, we also evaluated anxiety-related behavior using the HB test, as shown in Table 7. For the young and adult IR8d animals, the number of head dips was significantly lower ($P < 0.01$ for adult and $P < 0.05$ for young animals) than that for CTL animals, while the number of leg dips was significantly higher ($P < 0.01$) (Table 7). Young IR15d animals showed no alterations in the HB test when compared to CTL animals, but the numbers of head and leg dips in adult IR15d animals were significantly higher ($P < 0.05$) than in control animals (Table 7).

# Discussion

**Evolution of F1 generation body weight**

In our study, body weights of irradiated and control F1 were not statistically different until the postnatal age 70; however, from that age decreased relative to controls. This is interesting because fetal programming due to stress response, such as undernutrition, results in lower birth weight because the metabolic capacity in the uterus decrease, followed by a catch up growth and increase weight gain as adults as a compensatory response [5, 41]. Our data, however, suggest that LDIR induces a different outcome as the body weight decrease happens only at adult age. In several studies related to chronic mild stress (CMS), in which animals are exposed to a chronic and moderate stressful environmental condition, the body weight gain in adult life is reduced [42,43]. The rats of our study presented anxiety behavior and this characteristic body weight loss might be related to the animal's stress level. For ionizing radiation, Jensh *et al.* [44,45] researches using Wistar adult rats irradiated at doses ranging from 100 to 400 mGy on the GD9 or GD17 did not result in significant alterations in adult behavioral performance, and both Jensh's studies and our growth retardation persists postnatally.

**Newborn Reflexes and early motor development impairment**

Abnormalities on the central nervous system (CNS) appear to be diverse on LDIR studies, Jensh and Brent [44] neurophysiologic development study using Wistar rats irradiated with 200 mGy dose on the GD17 (stage of fetal development) dose reported no significant difference in eye-opening test. However, in our study IR8d (beginning of organogenesis phase) animals delayed the reflex response of eye-opening and the first climb on the negative geotaxis test compared to CTL animals, IR15d (ending of organogenesis phase) only delayed the first climb in relation to CTL. Our data indicate impairment in sensorimotor development and prenatal nervous system responsiveness when irradiation occurs in the organogenesis phase, especially for animals irradiated in the beginning of organogenesis (IR8d). The development of nervous system occurs in organogenesis phase.

**Behavioral disorders in exposed animals**

An anxious individual is not always anxious but is anxious more often than others, presenting disproportionate behavior to a stimulus, consequently, the best way to assess anxiety is by measuring how often and how intensely he exhibits anxious states [26]. In our study, through a sequence of tests, strong experimental evidence shows that even for 15 mGy exposure during pregnancy there is a decrease in central nervous system function. As expected, the effects of X-ray irradiation exposure in utero on rat development and behavior varied depending on gestational age at which irradiation occurred. In rodent studies [46], emotionality (or fear) and exploitation are inversely related since high emotionality inhibits exploration and low emotionality facilitates it. The results obtained from OF, HB and EPM tests are consistent in showing that there are deleterious effects related to the anxiety of the exposed animals.

The OF test results suggest that the prenatal LDIR modifies anxiety-related behaviors, especially in rats exposed during GD8. In comparison to young CTL animals, locomotor activity and exploratory behavior were higher in young IR8d rats, indicating excitability. Young IR15d rats, on the other hand, presented lower exploratory behavior than young CTL rats. Both locomotor activity and exploration are expected to decrease with age, as older animals present lower levels of anxiety, as well as in the present study, and it has been reported by Minamisawa and Hirokaga [47] that mice exposed to 100 mGy of γ rays during 14$^{th}$ day of gestation results in behavioral effects, but these effects decrease at late adulthood, as irradiated adult rats presented less difference than irradiated young rats when compared to control. Adult IR8d animals presented lower locomotor activity when compared to young IR8d, which was expected. Interestingly they also presented a lower locomotor activity when compared to adult CTL, indicating a normo-anxious behavior. The exploratory behavior of adult IR8d animals was lower than in young IR8d animals but higher when compared to adult CTL animals. Adult IR15d animals presented the expected decrease in locomotor activity and exploration with age but also presented a decrease in ambulation frequency when compared to adult CTL.

Following OF tests, the EPM results also indicated that young IR8d animals presented excitability when compared to CTL group, while young IR15d did not present any statistical differences. As expected, adult animals had a decrease in locomotor activity and exploration, but also in EPM tests IR8d animals had a more pronounced drop in ambulance values and adult IR15d animals presented a lower number of entries in open arms, indicating a higher state of anxiety.

In the HB test head dip is correlated to exploratory activity and according to Adzu *et al.* [48], the decrease in exploratory activity indicated by lower numbers of head dips is a measurement of depression of central nervous system (CNS) activity, whereas higher numbers of head dips suggest stimulation of CNS activity. Leg dip is correlated with motor coordination, therefore animals with a higher number of leg dips have motor coordination impairment. In the present study, young and adult IR8d data suggested depression on CNS and motor coordination impairment; as animals less exploited the device while presented a higher number of leg dip than CTL animals. The young IR15d animals did not present statistical difference to control group, adult IR15d data could not confirm motor coordination impairment in relation to control group because there is an increased exploration that can lead to the greater number of leg dip. A further intriguing aspect was the number of animal falls on the EPM test from the open arms of the device showing clear evidence of the motor coordination impairment of irradiated animals, especially IR8d.

The use of a combination of tests allows a more accurate observation of the animal's anxiety. The OF, EPM and HB generated data support each other, suggesting IR8d had higher exploratory behavior and decreased general anxiety when tested at 30 days old, however, the opposite occurs when examined at 70 days old. No effects were observed in the IR15d when tested at 30 days old, with increased anxiety at 70 days old. The EPM and HB tests also demonstrated a motor coordination impairment in irradiated animals, especially for IR8d animals.

**Final considerations**

To the best of our knowledge there are no studies showing biological effects for doses below 100 mGy. According to Wang and Zhou [49], the threshold for behavioral deleterious effects can only be detected from 100 mGy to 300 mGy, based on their study using mice irradiated with beta radiation from tritiated water (HTO). It is worth mentioning that gestation day stages of rat and mice differ: pre-implantation occurs for the first 7 days of gestation in rats, and 5 days in mice, organogenesis in rats between days 8 and 15, in mice between 7 and 12 days, and fetal development in rats between days 16 and 22, in mice between 13 and 20 days of gestation [5].

The three major stages of development in rats can be described as preimplantation, occurring during the first seven days of gestation, organogenesis, between days 8 and 15 of gestation, and fetal development, between days 16 and 22 of gestation. Therefore, the altered responsiveness of the postnatal nervous system observed mainly in IR8d animals is due to the effect of LDIR at the onset of organogenesis, when differentiation leads a process that less-specialized cells become more-specialized through the expression of a specific set of genes, by a cell signaling cascades, in which the initial signal is amplified. According to the "Law of Bergonie and Tribondeau" [50] or "Radiosensitivity Law" the high vascularization and high cellular proliferation rate lead to greater radiosensitivity, also radiosensitivity is inversely proportional to the degree of cell differentiation and directly proportional to the number of divisions required for the cell to reach its shape, so organogenesis is the phase of embryonic development in which there is a greater probability of deleterious effects.

Animal experimentation that analyzes the effects of LDIR generally uses doses higher than those used in human diagnostic procedures [17], therefore, the present study corroborates the search for effects in low-doses. However, it is important to note that even for 15 mGy exposure during the GD8 effects were observed: reduction of the weight gain, motor coordination impairment and affected sense of exploration. We think that probably fetal programming occurred in the irradiated animals, the "development origins of health and disease" (DOHaD) concept suggests that epigenetic patterns could explain the occurrence of LDIR-induced fetal programming affecting the responsiveness of the postnatal nervous system in IR8d animals [51]. The effects observed in IR15d animals were mild in relation to IR8d animals when compared to CTL group but showed affected emotionality without changing animal's anxiety [52].

In the present study, two different stages (IR8d and IR15d) of development were used to analyze behavior effects caused by irradiation during pregnancy, and the results agree with those reported elsewhere regarding that the begging of fetal development is the more sensitivity stage. Baskar and Devi [53] showed that 350 mGy mice exposure during organogenesis and early fetal period (11.5, 12.5, 14.5 days old) can decrease significantly locomotion and exploratory behavior in adult mice and, on day 11.5, embryos presented the highest radiosensitivity. Early studies of Werboff et al. [54] presented a significant decrease in locomotor activity in rats aged 55 days old that were prenatally irradiated with 220 mGy X-rays exposure on gestational day 10. However, this effect was not detected in animals irradiated on the gestational day 15 that, oppositely, showed an activity level increased as compared to the non-irradiated animals. Werboff et al. [54], also demonstrated that irradiation of whole-body pregnant rat can generate depression of activity in offspring first climb to animals exposed to 220 mGy dose in late gestation (day 15). On the other hand, although there is a limited literature data on behavioral effects of prenatal irradiation of rats, it is suggested that nonsignificant effects were observed at lower doses of 50 mGy [5,17,18].

Our results present a piece of evidence that even lower doses led to lower weight gain when adults, impairments in the central nervous system since infant phase, behavioral changes as activity and emotionality alters, persisting later into life and altered motor coordination. Animals irradiated early in organogenesis presented adverse and more severe effects on irradiated animals at the end of organogenesis, however, both groups had weight gain reduction and permanent behavioral changes evidencing fetal programming. We looked for following even lower doses than available data, as well as to use a considerable number of animals for a long time, in order to provide reliable data.

It is important to have more information about undesirable effects, further studies with low-doses are necessary to verify deleterious potential in human. Lack of information may lead to a misperception of the risk; therefore, an attempt should be made to increase and adopt effective radiation protection to low-doses, thus, information can help the radiologist to evaluate the best diagnostic procedure in a given clinical situation, ensuring the safety of the patient and the fetus.

# Acknowledgments

The authors wish to thank Dr. Luis Carlos Vulcano of the School of Veterinary Medicine and Animal Science at Sao Paulo State University for kindly providing the X-ray equipment. Rodrigo Sanchez Giarola is a recipient of a fellowship from *Coordenação de Aperfeiçoamento de Pessoal de Nível Superior* (Brazil [CAPES]; www.capes.gov.br – Finance code 001). The funding body had no role in the design, analysis, interpretation of data, or in writing the manuscript.

# Tables

## Table 1

Mean body weights of rats after and before parturition, litter size and pregnancy duration for irradiated (on gestational day 8 [IR8d] or 15 [IR15d]) and sham-irradiated control (CTL) rats. Values are mean ± SE, n=36.

|  | CTL (n=12) | IR8d (n=12) | IR15d (n=12) |
|---|---|---|---|
| Body weight F0 before parturition (g) | 321.7 ± 9.6 | 332.3 ± 13.5 | 342.1 ± 13.4 |
| Body weight F0 after parturition (g) | 260.4 ± 6.3 | 276.1 ± 8.7 | 266.6 ± 8.9 |
| Litter size | 10.1 ± 0.5 | 10.2 ± 0.7 | 10.9 ± 0.8 |
| Pregnancy duration (days) | 22.0 ± 0.1 | 21.7 ± 0.2 | 21.8 ± 0.3 |

## Table 2

Mean body weights of F1 irradiated (on gestational day 8 [IR8d] or 15 [IR15d]) and sham-irradiated control (CTL) rats at four different ages. Values are mean ± SE, n=144.

|  | CTL (n=48) | IR8d (n=48) | IR15d (n=48) |
|---|---|---|---|
| Body weight F1 pups 96 h after birth (g) | 12.2 ± 0.1 | 12.4 ± 0.1 | 12.4 ± 0.3 |
| Body weight F1 aged 30 days old (g) | 100.0 ± 2.2 | 102.2 ± 1.2 | 100.6 ± 1.5 |
| Body weight F1 aged 70 days old (g) | 383.8 ± 5.9 | **331.7 ± 2.0**\*\* | **355.1 ± 2.8**\*\* |
| Body weight F1 aged 91 days old (g) | 474.5 ± 4.8 | **412.5 ± 4.7**\*\* | **441.2 ± 2.6**\*\* |

*\*\*P < 0.01,* ANOVA followed by post-hoc Tukey's honestly significant difference test, compared with control (CTL).

## Table 3

Neurodevelopmental reflex tests in sham-irradiated control (CTL) and irradiated (on gestational day 8 [IR8d] or 15 [IR15d]) newborn rats: eye-opening, first climb and time to climb at 15 days old in negative geotaxis test. Values are mean ± SE, n=144.

| Treatment | CTL (n=48) | IR8d (n=48) | IR15d (n=48) |
|---|---|---|---|
| Eye-opening (days) | 12.8 ± 0.2 | **15.2 ± 0.7**\*\* | 12.3 ± 0.2 |
| First climb (days) | 9.4 ± 0.4 | **14.4 ± 0.5**\*\* | **12.2 ± 0.2**\*\* |
| Time to climb at 15 days old (s) | 5.4 ± 0.3 | 6.2 ± 0.3 | 5.9 ± 0.2 |

*\*\*P < 0.01,* ANOVA followed by post-hoc Tukey's honestly significant difference test, compared with control (CTL).

**Table 4**

Open field behavior of prenatally irradiated (on gestational day 8 [IR8d] or 15 [IR15d]) and sham-irradiated control (CTL) young (30 days old) and adult (70 days old) rats. Values given as means ± SE, n=144 (young) and n= 134 (adult).

|  | Young | | | Adult | | |
|---|---|---|---|---|---|---|
| Behavioral parameter | CTL (n=48) | IR8d (n=48) | IR15d (n=48) | CTL (n=47) | IR8d (n=40) | IR15d (n=47) |
| Number of squares entered | 87.6 ± 0.9 | **102.9 ± 1.4**\*\* | 83.8 ± 1.6 | 76.2 ± 1.7 | **54.9 ± 2.5**\*\* | **65.8 ± 0.9**\*\* |
| Rearing | 5.3 ± 0.1 | **7.6 ± 0.1**\*\* | **3.1 ± 0.2**\*\* | 3.3 ± 0.3 | **6.5 ± 0.4**\*\* | 2.7 ± 0.3 |
| Time to first cross the center | 163.1 ± 5.9 | 172.4 ± 4.6 | 166.8 ± 3.9 | 176.8 ± 1.6 | 175.4 ± 2.2 | 175.9 ± 2.5 |
| Number of crossings by the center | 0.7 ± 0.2 | 0.6 ± 0.1 | 0.8 ± 0.1 | 0.8 ± 0.3 | 0.8 ± 0.2 | 0.6 ± 0.1 |

*\*\*P < 0.01,* ANOVA followed by post-hoc Tukey's honestly significant difference test, compared with control (CTL).

**Table 5**

The behavior of young and adult rats in the elevated plus-maze test prenatally irradiated (on gestational day 8 [IR8d] or 15 [ IR15d] and sham-irradiated control (CTL) young (30 days old) and adult (70 days old) rats. Values given as means ± SE, n=144 (young) and n= 134 (adult).

|  |  | Young | | | Adult | | |
|---|---|---|---|---|---|---|---|
| Treatment | | CTL (n=48) | IR8d (n=48) | IR15d (n=48) | CTL (n=47) | IR8d (n=40) | IR15d (n=47) |
| Open arms | Number of entries | 12.4 ± 0.5 | **17.7 ± 0.5**\*\* | 12.6 ± 0.3 | 10.5 ± 0.3 | **7.6 ± 0.4**\*\* | **6.3 ± 0.4**\*\* |
| | Time spent (s) | 121.3 ± 3.5 | **152.3 ± 2.3**\*\* | 122.7 ± 1.7 | 102.4 ± 2.1 | **183.2 ± 3.6**\*\* | 92.3 ± 3.9 |
| Closed arms | Number of entries | 11.2 ± 0.3 | 12.1 ± 0.3 | 11.2 ± 0.2 | 7.5 ± 0.3 | 6.8 ± 0.3 | 8.3 ± 0.3 |
| | Time spent (s) | 113.5 ± 5.3 | **80.8 ± 1.9**\*\* | 120.7 ± 1.8 | 118.2 ± 2.9 | **94.5 ± 3.0**\*\* | 120.1 ± 3.1 |

*\*\*P < 0.01,* ANOVA followed by post-hoc Tukey's honestly significant difference test, compared with control (CTL).

**Table 6**

The percentages of young and adult rats in the elevated plus-maze test prenatally irradiated (on gestational day 8 [IR8d] or 15 [ IR15d] and sham-irradiated control (CTL) young (30 days old) and adult (70 days old) rats. Values given as means, n=144 (young) and n= 134 (adult).

|  | Young | Adult |
|---|---|---|

|                         | CTL (n=48) | IR8d (n=48) | IR15d (n=48) | CTL (n=47) | IR8d (n=40) | IR15d (n=47) |
|-------------------------|------------|-------------|--------------|------------|-------------|--------------|
| Treatment               |            |             |              |            |             |              |
| % time on open arms     | 51.7       | 65.3        | 50.4         | 46.4       | 66.0        | 43.5         |
| % open arms entries     | 52.5       | 59.4        | 52.9         | 58.3       | 52.8        | 43.2         |
| Total number of entries | 23.6       | 29.8        | 23.8         | 18.0       | 14.4        | 14.6         |

**Table 7**

The behavior of young and adult rats in the Hole board test prenatally irradiated (on gestation day 8 [IR8d] or 15 [IR15d]) and sham-irradiated control (CTL) young (30 days old) and adult (70 days old) rats. Values given as means ± SE, n=144 (young) and n= 134 (adult).

|       | Behavioral parameter | CTL (n=48)    | IR8d (n=48)     | IR15d (n=48)   |
|-------|----------------------|---------------|-----------------|----------------|
| Young | Number of head dip   | 15.5 ± 1.2    | **8.8 ± 0.3**\*\*  | 13.4 ± 0.4     |
|       | Number of legs dip   | 8.0 ± 0.4     | **11.4 ± 0.5**\*\* | 9.4 ± 0.4      |
|       | Behavioral parameter | CTL (n=47)    | IR8d (n=40)     | IR15d (n=47)   |
| Adult | Number of head dip   | 4.2 ± 0.4     | **2.6 ± 0.2**\*    | **5.6 ± 0.4**\*   |
|       | Number of legs dip   | 1.8 ± 0.2     | **3.7 ± 0.3**\*\*  | **2.6 ± 0.2**\*   |

*$P < 0.05$, ANOVA followed by post-hoc Tukey's honestly significant difference test, compared with control (CTL).

**$P < 0.01$, ANOVA followed by post-hoc Tukey's honestly significant difference test, compared with control (CTL).

# Additional Information

Author Contributions

R.S.G. and H.C.D. designed the project; R.S.G., G.H.O.A., T.H.H. and I.R.P. performed the housing, husbandry, weighing, behavioral tests, and video evaluation, under supervision of H.C.D. and A.F.G.. H.C.D. and A.F.G. contributed with lab equipment, test devices and animals. R.S.G., N.L., L.V., H.C.D. and J.M. performed data analysis. R.S.G. and M.L.A. wrote the manuscript. All authors read and approved the final manuscript.

Competing Interests

The authors declare that they have no competing financial and non-financial interests.

# Data Availability Statement

Authors commit to make all data available.